\documentclass[journal,twoside,web]{ieeecolor}
\usepackage{tmi}
\usepackage{cite}
\usepackage{amsmath,amssymb,amsfonts}
\usepackage{graphicx}
\usepackage{textcomp}
\usepackage{subcaption}
\usepackage{bm}
\usepackage{float}
\usepackage{dcolumn}
\usepackage{url}
\usepackage{booktabs}

\usepackage[
     breaklinks = true,
     colorlinks = true,
     linkcolor = blue, 
     urlcolor  = black, 
     citecolor = blue,
     anchorcolor = green,
     ]{hyperref}
\usepackage{algcompatible, algorithm, setspace}

\newcommand{\bA}{\mathbf{A}}
\newcommand{\bB}{\mathbf{B}}

\newcommand{\bC}{\mathbf{C}}
\newcommand{\bD}{\mathbf{D}}

\newcommand{\bR}{\mathbf{R}}

\newcommand{\bI}{\mathbf{I}}

\newcommand{\bsm}{\boldsymbol{m}}

\newcommand{\bsv}{\boldsymbol{v}}

\newcommand{\bsx}{\boldsymbol{x}}

\newcommand{\bsy}{\boldsymbol{y}}
\newcommand{\bsY}{\boldsymbol{Y}}

\newcommand{\bspi}{\boldsymbol{\pi}}
\newcommand{\bsmu}{\boldsymbol{\mu}}

\newcommand{\bsalpha}{\boldsymbol{\alpha}}
\newcommand{\bsbeta}{\boldsymbol{\beta}}

\newcommand{\bsSigma}{\boldsymbol{\Sigma}}

\newcommand{\bstheta}{\boldsymbol{\theta}}

\newcommand{\E}{\mathbb{E}}

\newcommand{\Pro}{\mathbb{P}}
\newcommand{\R}{\mathbb{R}}

\usepackage{multicol}
\def\BibTeX{{\rm B\kern-.05em{\sc i\kern-.025em b}\kern-.08em
    T\kern-.1667em\lower.7ex\hbox{E}\kern-.125emX}}
\begin{document}
\title{
Spectral image clustering on dual-energy CT scans using functional regression mixtures
}
\author{Segolene Brivet, Faicel Chamroukhi, Mark Coates \IEEEmembership{Senior Member, IEEE}, Reza Forghani, and Peter Savadjiev 
\thanks{
This research is supported by Fonds de recherche du Québec - Santé (FRQS) and Fondation de l’association des radiologistes du Québec (FARQ) (R.F.); by the Natural Sciences and Engineering Research Council of Canada (NSERC), [funding reference number 260250] (M.C.); and by the French National Research Agency (ANR), [grant SMILES ANR-18-CE40-0014] (F.C.).}
\thanks{S. Brivet and M. Coates are in the Electrical and Computer Engineering Department, McGill University, Montreal, QC H3A 0G4, Canada (e-mail: segolene.brivet@mail.mcgill.ca; mark.coates@mcgill.ca).}
\thanks{F. Chamroukhi is with Normandie Univ, UNICAEN, CNRS, LMNO, 14000 Caen, France (e-mail: faicel.chamroukhi@unicaen.fr).}
\thanks{R. Forghani and P. Savadjiev are in the Augmented Intelligence and Precision Health Laboratory (AIPHL), Department of Radiology, McGill University, Montreal, QC H3G 1A4, Canada (reza.forghani@mcgill.ca; peter.savadjiev@mcgill.ca)}
}

\maketitle

\begin{abstract}  

  Dual-energy computed tomography (DECT) is an advanced CT scanning technique enabling material characterization not possible with conventional CT scans. It allows the reconstruction of energy decay curves at each 3D image voxel, representing varying image attenuation at different effective scanning energy levels. In this paper, we develop novel functional data analysis (FDA) techniques and adapt them to the analysis of DECT decay curves. More specifically, we construct functional mixture models that integrate spatial context in mixture weights, with mixture component densities being constructed upon the energy decay curves as functional observations. We design unsupervised clustering algorithms by developing dedicated expectation maximization (EM) algorithms for the maximum likelihood estimation of the model parameters.
  To our knowledge, this is the first article to adapt statistical FDA tools and model-based clustering to take advantage of the full spectral information provided by DECT. 
  We evaluate our methods on 91 head and neck cancer DECT scans. We compare our unsupervised clustering results to tumor contours traced manually by radiologists, as well as to several baseline algorithms. Given the inter-rater variability even among experts at delineating head and neck tumors, and given the potential importance of tissue reactions surrounding the tumor itself, our proposed methodology has the potential to add value in downstream machine learning applications for clinical outcome prediction based on DECT data in head and neck cancer.

\end{abstract}




\section{Introduction}
\label{sec:introduction}

Computed Tomography (CT) has been one of the most common and widespread imaging techniques used in the clinic for the last few decades. There is increasing interest in a more advanced CT technique known as dual-energy CT (DECT) or spectral CT that enables additional material or tissue characterization beyond what is possible with conventional CT. In 
particular, DECT makes possible the computation of image attenuation levels at multiple effective energy levels. 
This results in the association of a decay curve with each reconstructed image voxel, representing energy-dependent changes in attenuation at that body location.


In this paper, we evaluate the application of DECT to head and neck squamous cell carcinoma (HNSCC). Current image-based evaluation of HNSCC tumors in clinical practice is largely qualitative, based on a visual assessment of tumor anatomic extent and basic one- or two-dimensional tumor size measurements. However, the frequently complex shape of mucosal head and neck cancers and at times poorly defined boundaries and potential adjacent tissue reactions can result in a high inter-observer variability in defining the extent of the tumor~\cite{Vinod2016, Hong2004, Veen2019, GUDI2017}, especially among radiologists without sub-specialty expertise in head and neck imaging. Furthermore, there is strong evidence that alterations of gene expression and protein-protein interactions in the peri-tumoral tissue or normal adjacent tissue may play a critical role in the evolution and risk of recurrence of HNSCC tumors (e.g., \cite{ImpactGanci2017,ImpactAran2017}). Yet these adjacent regions may not be obviously salient upon visual image examination.
For all these reasons, the accurate and consistent determination of a predictive region around the tumor is essential, both for conventional staging which determines patient management and for future automated quantitative image-based predictive algorithms based on machine learning. Such a predictive region may include the tumor itself, but also  surrounding tissues of biological relevance.

With these considerations in mind, DECT provides new and unexplored opportunities to answer an important question: what can be learned from DECT about tumor heterogeneity and its associations with surrounding tissues? As a first step towards answering this question, in this paper, we adapt functional data analysis techniques to DECT data in order to explore underlying patterns of association in and around the tumor. Functional data analysis (FDA) is a classical branch of statistics dedicated to the analysis of functional data, in situations where each data object is considered a function. This is particularly appropriate for DECT image data, as each 3D image voxel is associated with a curve of image intensity decay over multiple reconstructed energy levels (more details are provided below in Section \ref{subsec:DECT-data}). Thus, we adapt FDA statistical models for the clustering of 3D image voxels based on the full functional information provided by the decay curves associated with each voxel.  More specifically, the architecture underpinning our proposed method is a functional mixture model, where the mixture component densities are built upon functional approximation of the spectral decay curves at each image voxel, and the mixture weights are constructed to integrate spatial constraints. We then derive an Expectation-Maximization (EM) algorithm to perform the maximum likelihood estimators of the model parameters.

To our knowledge, this is the first article to propose spatial clustering utilizing the full spectral information available in DECT data, based on an appropriate FDA statistical framework. Existing methods for automatic tumor delineation in DECT (reviewed in detail in Section \ref{subsec:LitReviewSegm}) are mostly based on deep learning techniques and utilize only a small subset of the available information, due to the sheer amount of 4D (spatial + spectral) data available in a single DECT scan.

We apply the proposed methodology on 91 DECT scans of HNSCC tumors and we compare our results to manually traced tumor contours performed by an experienced expert radiologist. We also compare to other baseline clustering methods. However, tumor segmentation on its own is not a clinical outcome. A full demonstration of the clinical utility of our method necessitates an analysis of its ability to predict actual clinical outcomes, and how this prediction performance compares to the performance in the case of manually drawn contours, or contours drawn using alternative automatic methods. We leave this prediction analysis for a subsequent paper. As the first article to adapt FDA statistical tools to DECT data, the main focus of the present paper is on the statistical methodology and on algorithm development. As such, we can summarize our contributions as follows:
\begin{enumerate}
    \item we extend the statistical framework of functional mixture models to the spatio-spectral heterogeneous DECT data. In particular, DECT energy decay curves observed at each image voxel are modeled as spatially-distributed functional observations.  
    \item the proposed model is a novel generative mixture density model,
      with an efficient EM learning algorithm.
    \item we develop an unsupervised clustering of voxels incorporating  full spectral information from DECT data.
    \item to our knowledge, this is the first time that FDA statistical tools are applied to the analysis of DECT data.
\end{enumerate}
The source codes with summarising pseudo-codes for the proposed methods are publicly available \footnote{\url{https://github.com/segobrivet/DECT_clustering}}, free of charge.

\section{Background and related work}    \label{sec:background}

\subsection{Dual-energy CT}    \label{subsec:DECT-data}

\begin{figure}[!t]
\begin{center}
\begin{subfigure}[t]{.4\textwidth}
  \centering
  \includegraphics[trim={0 0 0 0},clip, width=\linewidth]{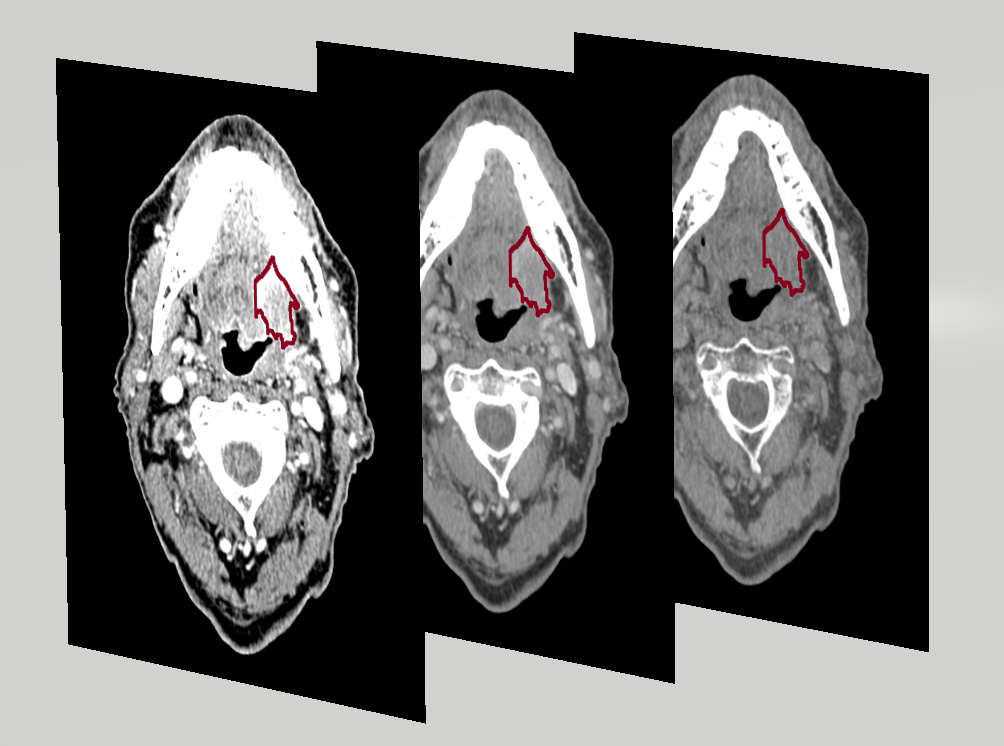}  
  \caption{\textbf{2D slices of VMIs at 40,65,140keV with tumor contour in red.} \textit{At lower energy levels, VMIs are more constrasted; at higher levels, VMIs are less noisy. A VMI at 65keV is 
  similar to a standard CT scan.}}
\label{fig:DECT-VMIs}
\end{subfigure}
\begin{subfigure}[t]{.45\textwidth}
  \centering
  \includegraphics[trim={20 13 30 0},clip, width=\linewidth]{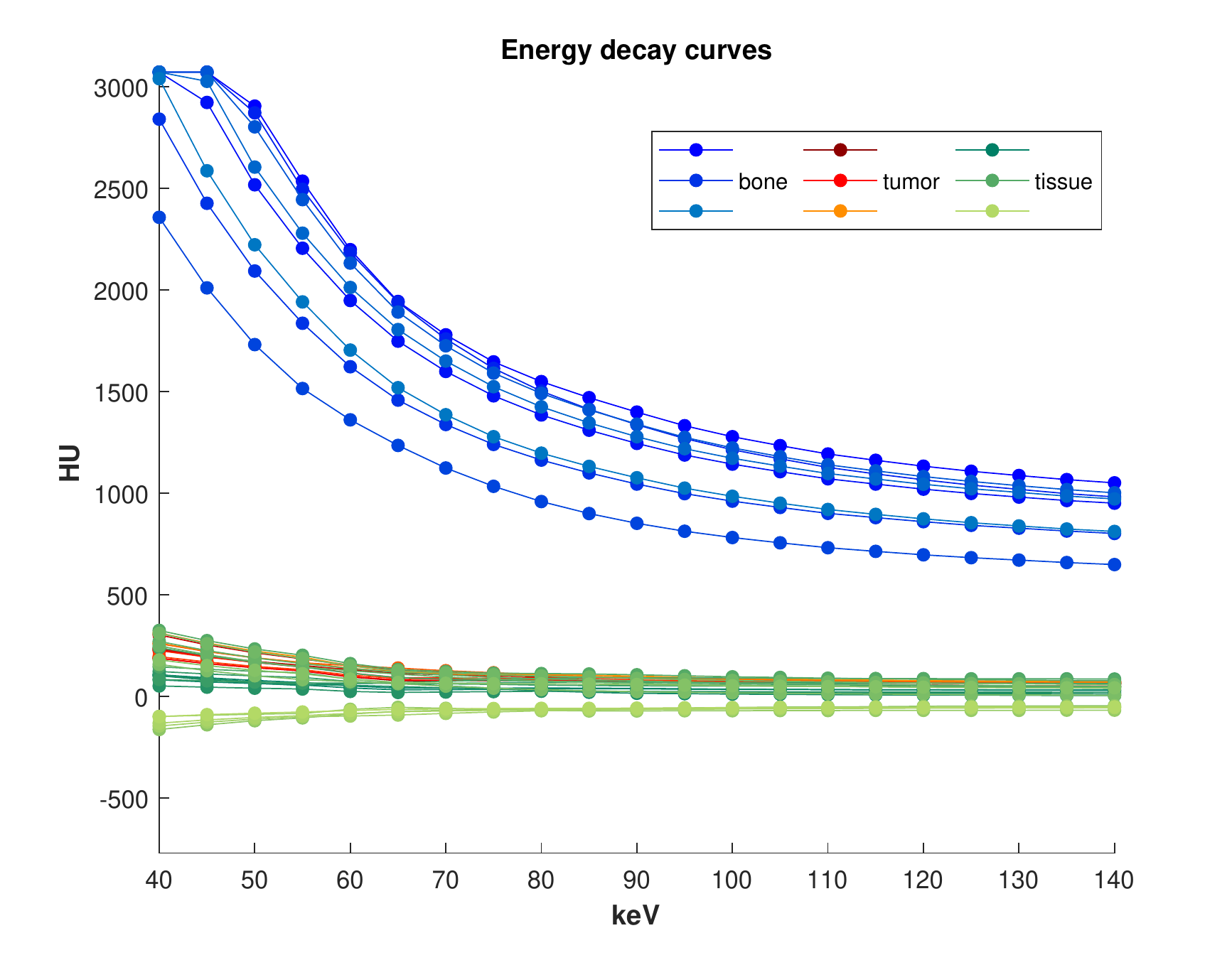} 
  \caption{\textbf{Examples of decay curves for different body locations.} \textit{A blue (resp. red, green) curve represents attenuation information stored at one voxel within bone (resp. tumor, tissue).}}
\label{fig:DECT-decay-curves}
\end{subfigure}
\end{center}
\end{figure}

DECT data may be viewed as a 4D image of a patient: a 3D body volume over a range of energy levels. The dual-energy image acquisition using two x-ray energy peaks at the source provides enough attenuation information to be combined and to be able to reconstruct a curve at multiple ``virtual monochromatic" energy levels. These simulate what the attenuation (in Hounsfield units; HU) would be if the study was acquired with a monochromatic x-ray beam at that energy value (in kilo-electron-Volt; keV). The reconstructed curve of attenuation numbers over each energy level translates the energy-dependent changes and is commonly called the spectral Hounsfield unit attenuation curve, or an \textit{energy decay curve}~\cite{intro13}. In our method, we will make use of this spectral information through functional approximations, and thus consider the curves as functional observations.
An energy decay curve has been calculated for each image voxel, thus a DECT scan is represented as a 4D image with 3 dimensions for X, Y and Z spatial coordinates and 1 dimension for energy level coordinates. The virtual monochromatic image (VMI) is the 3D image representation at a given energy level. See Fig.~\ref{fig:DECT-VMIs} for examples of a 2D slice from different VMIs and Fig.~\ref{fig:DECT-decay-curves} for examples of decay curves for different tissue characteristics.

\subsection{Segmentation of dual-energy CT data}    \label{subsec:LitReviewSegm}

Segmentation is a process of delineating an image region of interest. For example, radiation oncologists usually manually segment tumors for radiation planning. Automatic tumor segmentation has a long history of developments: from knowledge-driven early techniques to data-driven newer techniques, algorithms aims to extract image features to make a decision on region boundaries~\cite{PeterKnowledgeDriven2019}. However this process remains challenging in medical imaging due to the heterogeneity over the image or the acquisition process, most of current algorithms need manual adjustments on the result~\cite{PeterBiomarkers2019}.

In head and neck CT imaging, the difficulty to contour precisely a tumor region results in large inter-observer variability in segmentation results, even among trained radiation oncologists.
 A study among radiologists from 14 different institutions obtained a median Dice Similarity Score (DSC) ranging from 0.51 to 0.82 \cite{Veen2019}, depending on the delineation criteria used. Another study assessing the same variability among 3 experienced radiologists over 10 tumors obtained a mean DSC of 0.57 \cite{GUDI2017}.

To the best of our knowledge, only a few studies have focused on DECT segmentation. These employ deep learning approaches \cite{DECTorganSegm, DECTorganSegm2, DECTorganSegmWang}. The four dimensions of the data required workarounds in order to apply neural networks. For example, using two VMIs sampled from the energy level spectrum, one at a low- and one at a high-energy level, Chen et al. in ~\cite{DECTorganSegm} merge the two VMIs in a layer connected to a U-Net architecture \cite{Unet}.  Wang et al. in \cite{DECTorganSegmWang} learn features from two pyramid networks on the two VMIs independently and combine them through deep attention into a mask scoring regional convolutional neural network (R-CNN). They achieved good performance in segmenting large-sized organs (DSC larger than 0.8); and performance is less impressive for small-sized organs (DSC between 0.5 and 0.8).

\subsection{Decay curve clustering via functional data analysis}

FDA aims to represent infinite-dimensional functional data into a finite-dimensional vector of coefficients~\cite{FDAreview}.
To achieve this, FDA consists in expanding functional data into function bases.
One approach relies on projection on bases which consists in projecting functional data onto finite dimensional function bases (e.g. splines, B-splines, polynomials, Fourier, wavelet). It associates a finite vector of projection coefficients. This is what we use in this paper.
Analogously another common approach would be to run a functional principal component analysis (fPCA) to get a basis of eigenfunctions of the covariance of the process describing our functional data. It associates a (truncated) projection vector of PCA coefficients.

Our objective is to partition our data, modeled with FDA, in different groups of voxels having similar decay curve characteristics. Among the available clustering approaches (e.g. centroid-based clustering such as k-means, connectivity-based clustering such as hierarchical clustering, density-based clustering such as DBSCAN~\cite{DBSCAN1996}, distribution-based clustering with model-based methods), since we have a model for each decay curve, a model-based approach is preferred. 

Model-based clustering is a thoroughly developed field \cite{mclachlan_basford88,Fraley2002-MBC}, in particular for multivariate analysis. 
Model-based clustering approaches rely on the finite mixture modeling framework \cite{McLachlan2000FMM} to represent the density of a set of independent multivariate observations and on an optimization algorithm to automatically find a partition into groups of such observations. 

To represent different groups of data, mixture models assume each datum to follow a known distribution (e.g. Gaussian), and build a mean representation (i.e. model) for each group of data.
The mixture model calculates, for each data point, a value defined by the sum over $k=\{1...\#\text{groups}\}$, of the probability distribution function (pdf) that this point belongs to group $k$ model, emphasized by a weight giving a higher or lower chance of belonging to this group (derived in Section~\ref{subsec: gen-mdl}).
Mixture models have the advantage of being interpretable, parametric thus well-understood, and flexible as the pdf  modeling the data in each cluster is chosen explicitely. 

The Expectation-Maximization (EM) algorithm~\cite{Dempster1977} is a popular and adapted tool with desirable properties that can be used to conduct an iterative estimation of the model parameters and thus the cluster membership probabilities. 

Mixture models for clustering have been applied and adapted to different kind of data, including time-series data~\cite{Same2011}, gene expression data~\cite{YeungMBC2001}, or 3D noisy medical images~\cite{Balafar2011}, spatio-temporal data (non image data)~\cite{SpatioSpectral2021}.
They also have been recently investigated for functional data~\cite{Chamroukhi-FDA-2019}, and thus provide an avenue to model the spectral decay curves, but in this context of spectral images, we also need to incorporate the spatial information into the clustering. 

A related idea was proposed in \cite{Tino2014} to develop a spatio-temporal mixture of Hidden Process Models for fMRI analysis. The authors built a temporal probabilistic model, and reshaped the prior probability with spatial constraints to determine a ``region of influence'' for the temporal model. A specification of our model covers this approach, and our model go further in generalizing it via the construction of more flexible Gaussian-mixture weights around the spatial coordinates. The resulting model enjoys better numerical learning properties with faster convergence due to closed-form updating rules, for the spatial weights parameters. Alternative constructions of the proposed mixture model are also presented in order to validate the method and to accommodate potential user specifications.

\section{Methodology}  \label{sec: methodology}
\subsection{Generative modeling framework}    \label{subsec: gen-mdl}
We adopt the framework of generative modeling for image clustering using 
different families of extended mixture distributions.  
The general form of the generative model for the image
assumes that the $i$th datum (i.e., pixel, voxel) in the image has the general semi-parametric mixture density:
\begin{equation}
    f_i(\bstheta) = \sum_{k=1}^K \pi_{ik} f_i(\bstheta_k)\,,
    \label{eq:general image mixture model}
\end{equation}
which is a convex combination of $K$ component densities, $f_i(\bstheta_k)$, $k\in[K]=\{1,\cdots,K\}$, weighted by non-negative mixture weights $\pi_{ik}$ that 
sum to one, that is $\sum_{k=1}^K\pi_{ik}=1$ for all $i$, $i\in[n]$. The unknown parameter vector $\bstheta$ of density \eqref{eq:general image mixture model} is
composed of the set of component density parameters  $\{\bstheta_k\}$ and their associated weights $\{\pi_{ik}\}$, i.e., $\bstheta =\{\pi_{ik},\bstheta_k\}_{k=1}^K$.

From the perspective of model-based clustering of the image, each component density can be associated with a cluster, and hence the clustering problem becomes one of parametric density estimation. Suppose that the image has $K$ segments and let $Z_i\in[K]$ be the random variable representing the unknown segment label of the $i$th observation in the image. Suppose that the distribution of the data within each segment $k\in[K]$ is $f(\bstheta_k)$, i.e, $f_{i|Z_i=k}(\bstheta) =f_i(\bstheta_k)$. 
Then, from a generative point of view, model \eqref{eq:general image mixture model} is equivalent to
 {\it i)} sampling a segment label according to the discrete distribution with parameters being the mixture weights $\bspi = \{\pi_{1},\cdots,\pi_{K}\}$,
  then 
   {\it ii)} sampling an observation $\text{Im}_i$ from the conditional distribution $f(\bstheta_k)$.
 Given a model of the form $\eqref{eq:general image mixture model}$ represented by $\widehat\bstheta$, 
 typically fitted by maximum likelihood estimation (MLE) from the $n$ observations composing the image $\text{Im}_n$,  as
 \begin{equation}
    \widehat \bstheta \in \arg\max_{\bstheta\in \Theta}\log L(\bstheta|\text{Im}_n)
\end{equation}
where 
$L(\bstheta|\text{Im}_n)$ is the likelihood of $\bstheta$ given the image data $\text{Im}_n$, 
 then, the segment labels 
 can be determined via 
 the Bayes' allocation rule,
 \begin{equation}
 \widehat{Z}_i = \arg\max_{k\in [K]}\Pro(Z_{i}=k|\text{Im}(i);\widehat \bstheta)\,,
      \label{eq:General Bayes Rule}
\end{equation}
which consists of maximizing the conditional probabilities  \begin{equation}
     \Pro(Z_{i}=k|\text{Im}(i);\widehat \bstheta) = \frac{\widehat\pi_{ik}f_i(\widehat\bstheta_k)}{f_i(\widehat\bstheta)}\cdot
     \label{eq:General Conditional Probs}
 \end{equation}
 that the $i$th observation originates from segment $k$, $k\in[K]$, given the image data and the fitted model.

Model \eqref{eq:general image mixture model} has  many different specifications in the literature depending on the nature of the data generative process, resulting in a multitude of choices for the mixture weights and for the component densities. 
 {Mixtures of multivariate distributions} \cite{McLachlan2000FMM}
 are in particular more popular in model-based clustering of vectorial data using multivariate mixtures. These include multivariate Gaussian mixtures \cite{Fraley2002-MBC,mclachlan_basford88},
where $\pi_{ik} = \pi_k$, $\forall i$, are constant mixture weights, and the component densities $f_i(\bstheta_k) = \phi_i(\bsmu_k,\bsSigma_k)$ are multivariate Gaussians with means $\bsmu_k$ and covariance matrices $\bsSigma_k$.
Mixtures of regressions models, introduced in \cite{MixOfReg}, are common in modeling and clustering of regression-type data. For example, in the widely used Gaussian regression mixture model \cite{GaussRegMixModels}, we have constant mixing proportions, i.e. $\pi_{ik} = \pi_k$, 
and the mixture components $f_i(\bstheta_k)$'s are Gaussian regressors 
$\phi(\cdot, \bsbeta_k^\top\bsx_i,\sigma_k^2)$
 with typically linear means $\bsbeta_k^\top\bsx_i$ and variances 
$\sigma_k^2$, in the case of univariate response.

In this paper, we consider a more flexible mixture of regressions model in which both the mixture weights and the mixture components are covariate-dependent, and are constructed upon flexible semi-parametric functions.
 More specifically, in this full conditional mixture model, the mixture weights $\pi_{ik}$ are constructed upon parametric functions $\pi_{ik} = \pi_k(\cdot, \bsx_i;\bsalpha)$ of some covariates $\bsx_i$'s represented by a parameter vector $\bsalpha$, and 
the regression functions  $f_i(\bstheta_k)$ are  Gaussian regressors $\phi(\cdot, \mu(\bsx_i;\bsbeta_k),\sigma_k^2)$
 with semi-parametric (non-)linear mean functions $\mu(\bsx_i;\bsbeta_k)$.  
This flexible modeling allows us to better capture more non-linear relationships in the functional data via the semi-parametric mean functions. Heterogeneity is accommodated via the mixture distribution, and spatial organization can be captured  via spatial-dependent mixture weights.

\subsection{Spatial mixture of functional regressions for dual-energy CT images}
We propose a spatialized mixture of functional regressions model, adapted to the given type of image data, for the model-based clustering of dual-energy CT scans. 
The images we analyse include spectral curves for each 3D-voxel. Each image, denoted $\text{Im}$, is represented as a sample of $n$ observations,
$\text{Im} = \{\bsv_i, \bsx_i, \bsy_i\}_{i=1}^n$
where $\bsv_i = (v_{i1},v_{i2},v_{i3})$ is the $i$th voxel 3D-spatial coordinates.
The $i$th voxel is represented by the  curve $(\bsx_i,\bsy_i)$ composed of HU attenuation values
 $\bsy_i = (y_{i1},\ldots,y_{im})$  measured at energy levels (covariates) $\bsx_i = (x_{i1},\ldots,x_{im})$,	with $m$ being the  number of energy levels.

To accommodate the spatial organization of the image together with the functional nature of each of its voxels, 
we propose spatialized conditional extensions of the general family of 
model \eqref{eq:general image mixture model}, in which we model the $i$th voxel observation  of the image using the conditional density $f(\bsy_i|\bsx_i,\bsv_i;\bstheta)$   that relates the attenuation curve levels $\bsy_i$ given the associated energy levels $\bsx_i$, and spatial location $\bsv_i$, via a convex combination of (non-)linear (semi-)parametric functional regressors $f(\bsy_i|\bsx_i; \bstheta_k)$
with spatial weights $\pi_k(\bsv_i;\bsalpha)$, that is:
\begin{equation}
    f(\bsy_i|\bsx_i,\bsv_i;\bstheta) = \sum_{k=1}^K \pi_k(\bsv_i;\bsalpha) f(\bsy_i|\bsx_i; \bstheta_k)\,.
    \label{eq:general spatial fMoE}
\end{equation}
To this purpose, we consider two different spatial constructions of the mixing weights (gating functions) $\pi_k(\bsv_i;\bsalpha)$: (i) softmax gates; and (ii) normalized Gaussian gates. The latter is an appropriate choice if 
more approximation quality is needed, 
and facilitates the computations in the learning process. 
We also consider different families to model the functional regressors, including spline and B-spline regression functions that enjoy better curve approximation capabilities, compared to linear or polynomial regression functions.

\subsubsection{Functional regression components}
We have a 3D image volume over a range of energy levels that provide, for each voxel $i$, an attenuation curve $(\bsx_i,\bsy_i)$ which represents energy dependent changes in attenuation, which enables a better characterization of the tissue at voxel $i$. 
We therefore model the component densities $f(\bsy_i|\bsx_i; \bstheta_k)$ as functional regression models constructed upon the attenuation curves as functional observations. This allows us to accommodate the spectral curve nature of the data. 
More specifically, in the case of univariate energy levels, 
we use smooth functional approximations
to model, for the $i$th voxel, the mean spectral curve of the $k$th component $\mu(\bsx_i;\bsbeta_k) =\E_{\bstheta}[\bsY_i|Z_i=k,\bsx_i]$, that is  
$\mu(\bsx_i;\bsbeta_k)= (\mu(x_{i1};\bsbeta_k),\ldots,\mu(x_{im};\bsbeta_k))$ 
using polynomial or (B)-spline functions,
whose coefficients are $\bsbeta_k$'s. 


The conditional density model for each regression is then modeled as a functional Gaussian regressor defined by
$f(\bsy_i|\bsx_i; \bstheta_k)=\phi_{m}\!\left(\bsy_i; \mu(\bsx_i;\bsbeta_k), \sigma^2_k \bI\right)$, with $\mu(\bsx_i;\bsbeta_k) = \bB(\bsx_i)\bsbeta_k$ being the function approximation onto polynomial or (B-)spline  bases $\bB(\bsx_i)$,   
and the matrix form of the functional regression model 
is then given by
\begin{equation}
    f(\bsy_i|\bsx_i; \bstheta_k)=\phi_{m}(\bsy_i; \bB(\bsx_i)\bsbeta_k, \sigma^2_k \bI)\,,
\label{eq:functional regression functions}
\end{equation}
where $\bstheta_k = (\bsbeta^\top_k,\sigma_k^2)^\top\in \R^{p+q+2}$ is the unknown parameter vector of regression $k$. 

\subsubsection{Spatial gating functions}

The constructed functional mixture of regressions model \eqref{eq:general spatial fMoE} specifically integrates the spatial constraints in the mixture weights $\pi_{k}(\bsv_i;\bsalpha)$ via functions of the spatial locations $\bsv_i$ parametrized by vectors of coefficients $\bsalpha$.
We investigate two choices to this end. 
The first proposed model is a spatial softmax-gated functional mixture of regression and is defined by \eqref{eq:general spatial fMoE} with a softmax gating function:
\begin{equation}
\pi_{k}(\bsv_i;\bsalpha) = \frac{\exp{(\bsalpha^\top_k\bsv_i)}}{1+\sum_{k^\prime=1}^{K-1}\exp{(\bsalpha^\top_{k^\prime}\bsv_i)}}\,, 
\label{eq:Softmax Gating Net}
\end{equation}
where $\bsalpha = (\bsalpha^\top_1,\ldots,\bsalpha^\top_K)^\top$ is the unknown parameter vector of the gating functions. 
 We will refer to this model, defined by \eqref{eq:general spatial fMoE}, 
 \eqref{eq:functional regression functions} and \eqref{eq:Softmax Gating Net},
 as the Spatial Softmax-gated Mixture of Functional Regressions, abbreviated as {\bf SsMFR}. 
The softmax modeling of the mixture weights is a standard choice known in the mixtures-of-experts community. However, its optimization performed at the M-Step of the EM algorithm, is not analytic and requires numerical Newton-Raphson optimization. This can become costly, especially in larger image applications, such as the one we address.

In the second proposed model, we use
a spatial Gaussian-gated functional mixture of regressions, defined by \eqref{eq:general spatial fMoE} with a Gaussian-gated function:
\begin{equation}
    \pi_{k}(\bsv_i;\bsalpha) = \frac{w_k \phi_3(\bsv_{i};\bsmu_k,\bR_k)}{\sum_{\ell=1}^{K}w_\ell \phi_3(\bsv_{i};\bsmu_\ell,\bR_\ell)}\,,
\label{eq:Gaussian Gating Net}
\end{equation}
in which $w_k$ are non-negative weights that sum to one, 
$\phi_d(\bsv_{i};\bsmu_k,\bR_k)$ is the density function of a multivariate Gaussian  vector of dimension $d$ with mean $\bsmu_k$ and covariance matrix $\bsSigma_k$, and
$\bsalpha = (\bsalpha^\top_1,\ldots,\bsalpha^\top_K)^\top$ is the parameter vector of the gating functions 
with $\bsalpha_k = (w_k,\bsmu^\top_k,{\text{vech}(\bR_k)^\top})^\top$. 

We will refer to this model, defined by \eqref{eq:general spatial fMoE}, \eqref{eq:functional regression functions} and \eqref{eq:Gaussian Gating Net}, as the Spatial Gaussian-gated Mixture of Functional Regressions, abbreviated as {\bf SgMFR}.
This Gaussian gating function was introduced in~\cite{jordan-and-xu-1995} to bypass the need for an additional numerical optimization in the inner loop of the EM algorithm. We obtain a closed form updating formula at the M-Step, that is detailed in the next section presenting the derived EM algorithm.

\subsubsection{MLE of the SgMFR model via the EM algorithm}  \label{sec:MLE-SgMFR}
Based on equations~\eqref{eq:general spatial fMoE}, \eqref{eq:functional regression functions} and \eqref{eq:Gaussian Gating Net}, the SgMFR 
joint density
$f(\bsy_i,\bsx_i,\bsv_i;\bstheta)$
 is then derived 
and the joint log-likelihood we maximize via EM is 
{\small
\begin{align}\label{eq:joint loglik}
&\log L(\bstheta) = \sum_{i=1}^{n}\log f(\bsy_i,\bsx_i,\bsv_i;\bstheta) \nonumber \\
&    = \sum_{i=1}^{n}\log \sum_{k=1}^{K} w_k \phi_3(\bsv_{i};\bsmu_k,\bR_k)\phi_{m}(\bsy_i; \bB(\bsx_i)\bsbeta_k, \sigma^2_k \bI)\cdot
\end{align}
}
The complete-data log-likelihood, upon which the EM algorithm is constructed, is 
{\small
\begin{align}
\!\!\!    & \log L_c(\bstheta) = \nonumber \\
\!\!\!    & \sum_{i=1}^{n}\sum_{k=1}^{K}\! Z_{ik} \log\! \left[w_k \phi_3(\bsv_{i};\bsmu_k,\bR_k)
    \phi_{m}(\bsy_i; \bB(\bsx_i)\bsbeta_k, \sigma^2_k \bI)
    \right]\!,
    \label{eq:complete joint log-lik}
\end{align}
}where $Z_{ik}$ is an indicator  variable such that $Z_{ik}=1$ if $Z_i=k$ (i.e., if the $i$th pair $(\bsx_i,\bsy_i)$ is generated from the $k$th regression component) and $Z_{ik}=0$,  otherwise.
\color{black}
The EM algorithm, after starting with an initial solution $\bstheta^{(0)}$, alternates between the E- and the M- steps until convergence (when there is no longer a significant change in the log-likelihood~\eqref{eq:joint loglik}).
\paragraph{The E-step}
\label{ssec: E-step}
Compute the expected complete-data log-likelihood (\ref{eq:complete joint log-lik}), given  image $\text{Im}_n$ and the current estimate $\bstheta^{(t)}$:
{\small
\begin{align}
 &Q(\bstheta;\bstheta^{(t)}) = \E\left[L_c(\bstheta)|\text{Im}_n;\bstheta^{(t)}\right]\,,    \label{eq:Q-function}\\
 &=\sum_{i=1}^{n}\sum_{k=1}^{K}\tau_{ik}^{(t)} \log \left[\alpha_k \phi_3(\bsv_{i};\bsmu_k,\bR_k)
    \phi_{m}(\bsy_i; \bB(\bsx_i)\bsbeta_k, \sigma^2_k \bI)\right],
    \label{eq:Q-function}\nonumber
\end{align}
}where $\tau_{ik}^{(t)} = \Pro(Z_i=k|\bsy_i,\bsx_i,\bsv_i;\bstheta^{(t)})$ given by
{\small
\begin{equation}
     \tau_{ik}^{(t)} 
     = \frac{ w^{(t)}_k \phi_3(\bsv_{i};\bsmu_k^{(t)},\bR_k^{(t)}) 
    \phi_{m}(\bsy_i; \bB(\bsx_i)\bsbeta^{(t)}_k, {\sigma^2_k}^{(t)} \bI)
    }{f(\bsv_i, \bsx_i,\bsy_{i};\bstheta^{(t)})}
\label{eq:Conditonal Probs SgMFR}
\end{equation}}is the posterior probability that the observed pair $(\bsx_i, \bsy_i)$ is generated by  the $k$th regressor. This step therefore only requires  the computation of the posterior  component membership probabilities $\tau^{(t)}_{ik}$ $(i=1,\ldots,n)$, for $k=1,\ldots,K$.

\paragraph{The M-step}
\label{ssec: M-step EM}
Calculate the parameter vector update $\bstheta^{(t+1)}$ by maximizing the $Q$-function (\ref{eq:Q-function}), i.e.
$\bstheta^{(t+1)} = \arg \max_{\bstheta} Q(\bstheta;\bstheta^{(t)})$.
By decomposing the $Q-$function 
as 
\begin{equation}
\label{eq:Q-function decomposition}
Q(\bstheta;\bstheta^{(t)}) =\sum_{k=1}^{K}Q(\bsalpha_k;\bstheta^{(t)})+Q(\bstheta_k;\bstheta^{(t)})\,,
\end{equation}where 
{\small 
    $Q(\bsalpha_k;\bstheta^{(t)}) =  \sum_{i=1}^{n}\tau_{ik}^{(t)} \log \left[w_k \phi_3(\bsv_{i};\bsmu_k,\bR_k)\right]$ }
and \\
{\small
    $Q(\bstheta_k;\bstheta^{(t)}) = \sum_{i=1}^{n} \tau_{ik}^{(t)} \log \, [\phi_{m}(\bsy_i; \bB(\bsx_i)\bsbeta_k, \sigma^2_k \bI)]$}, 
the maximization can then be done by  $K$ separate maximizations w.r.t.\ the parameters of the gating  and the regression functions.

\noindent{\it Updating the gating functions parameters:} 
Maximizing~\eqref{eq:Q-function decomposition} w.r.t.\ $\bsalpha_k$'s corresponds to the M-Step of a Gaussian Mixture Model \cite{McLachlan2000FMM}. The closed-form expressions for updating the parameters are given by:
{\small
\begin{align}
 \!\!\!\!   w_{k}^{(t+1)} &= \sum_{i=1}^{n}\tau_{ik}^{(t)}\Big/n,
    \label{eq:MLE alphak}\\
 \!\!\!\!    \bsmu_{k}^{(t+1)} &=  \sum_{i=1}^{n}\tau_{ik}^{(t)}\bsv_i\Big/\sum_{i=1}^{n}\tau_{ik}^{(t)},
    \label{eq:MLE muk}\\
 \!\!\!\!    \bR_{k}^{(t+1)} &=  \sum_{i=1}^{n}\tau_{ik}^{(t)}(\bsv_i-\bsmu_k^{(t+1)})(\bsv_i-\bsmu_k^{(t+1)})^\top\!\!\Big/\!\sum_{i=1}^{n}\tau_{ik}^{(t)}\cdot
    \label{eq:MLE Rk}
\end{align}
}

\noindent{\it Updating the regression functions parameters:} Maximizing~\eqref{eq:Q-function decomposition} w.r.t.\ $\bstheta_k$'s corresponds to the M-Step of standard mixtures of experts with univariate Gaussian regressions. 
The closed-form updating formulas are given by:
{\small
\begin{align}
  \!\!\!\!   & \bsbeta^{(t+1)}_{k} \!\!=\!\! \Big[\sum_{i=1}^{n}  \tau_{ik}^{(t)} \bB^\top(\bsx_i)\bB(\bsx_i)\Big]^{-1} \sum_{i=1}^{n} \tau_{ik}^{(t)} \bB(\bsx_i)^\top\bsy_i\,,
     \label{eq:EM betak update}\\
  \!\!\!\!    & {\sigma_{k}^2}^{(t+1)} \!\!=\!\! \sum_{i=1}^{n}   \tau_{ik}^{(t)} (\bsy_i -  \bB(\bsx_i) \bsbeta_{k}^{(t+1)})^2\Big/\sum_{i=1}^n  \tau_{ik}^{(t)} m_i\,\cdot
    \label{eq:EM sigma2 update}
\end{align}
}


\subsection{Alternative two-fold approaches}
We also investigate an alternative approach to the one derived before, which consists of a two-fold approach, rather than simultaneous functional approximation and model estimation for segmentation. 
We first construct approximations of the functional data onto polynomial or (B-)splines bases  $\bB(\bsx_i)$
via MLE (ordinary least squares in this case) to obtain:
{\small
\begin{equation}
    \widehat\bsbeta_i = \Big[\bB(\bsx_i)^\top\bB(\bsx_i)\Big]^{-1} \bB(\bsx_i)^\top\bsy_i.
    \label{eq:OLS}
\end{equation}
}
Then, we model the density of the resulting coefficient vectors $\widehat\bsbeta_i$, which is regarded as the $i$th curve representative, by a mixture density with spatial weights of the form
{\small
\begin{equation}
f(\widehat\bsbeta_i, \bsv_i, \bstheta) =  \sum_{k=1}^{K} 
\pi_{k}(\bsv_i;\bsalpha) 
 \phi_{d}(\widehat\bsbeta_i; \bsm_k, \bC_k)\,,
\end{equation}}where $\bsm_k$ and $\bC_k$ are the mean and the covariance matrix of each component. 
The spatial weights $\pi_{k}(\bsv_i;\bsalpha)$ are normalized Gaussians as in \eqref{eq:Gaussian Gating Net} or softmax as in \eqref{eq:Softmax Gating Net}.
We will refer to these methods as Spatial Gaussian-gated (resp. softmax-gated) Mixtures of Vectorized Functional Regressions, {\bf SgMVFR} (resp. {\bf SsMVFR}).

The EM algorithm for fitting this mixture of spatial mixtures, constructed upon pre-computed polynomial or (B-)spline coefficients with its two variants for modeling the spatial weights, takes a similar form to the previously presented algorithm, and is summarized as follows.
The conditional memberships of the E-Step are given for the softmax-gated model by
{\small
\begin{equation}
\tau_{ik}^{(t)} 
= \frac{\pi_{k}(\bsv_i;\bsalpha^{(t)})\phi_{d}(\widehat\bsbeta_i; \bsm^{(t)}_k, \bC^{(t)}_k)}{f(\widehat\bsbeta_i|\bsv_i;\bstheta^{(t)})}\,,
    \label{eq:Condtional Probs SsGMR}
\end{equation}
}
and for the Gaussian-gated model by
{\small
 \begin{align}
 \tau_{ik}^{(t)} 
                &= \frac{ w^{(t)}_k \phi_3(\bsv_{i};\bsmu_k^{(t)},\bR_k^{(t)}) 
    \phi_{d}(\widehat\bsbeta_i; \bsm^{(t)}_k, \bC^{(t)}_k)
    }{f(\bsv_i, \widehat\bsbeta_i;\bstheta^{(t)})}\,.
    \label{eq:Condtional Probs SgGMR}
\end{align}
}
The latter has the same advantage as explained above. 
In the M-Step, the gating functions parameter updates are given by \eqref{eq:MLE alphak},\eqref{eq:MLE muk}, and \eqref{eq:MLE Rk} for the Gaussian-gated model, or through a Newton-Raphson optimization algorithm for the softmax-gated model.
The component parameter updates are those of classical multivariate Gaussian mixtures
{\small
\begin{align}
 \!\!\! \bsm_{k}^{(t+1)} &= \sum_{i=1}^{n}\tau_{ik}^{(t)}\widehat\bsbeta_i\Big/\!\sum_{i=1}^{n}\tau_{ik}^{(t)},
    \label{eq:MLE mk}\\
\!\!\!   \bC_{k}^{(t+1)} &= \sum_{i=1}^{n}\tau_{ik}^{(t)}(\widehat\bsbeta_i-\bsm_k^{(t+1)})(\widehat\bsbeta_i-\bsm_k^{(t+1)})^\top\!\Big/\!\sum_{i=1}^{n}\tau_{ik}^{(t)}\cdot
    \label{eq:MLE Ck}
\end{align}
}



In a nutshell, to compute a clustering of the image, 
the label of voxel $i$ given the fitted  parameters $\widehat\bstheta$ is calculated by the Bayes' allocation rule \eqref{eq:General Bayes Rule}, 
in which $\text{Im}(i)$ is the spatial coordinates of voxel $i$ with, either its direct spectral curve representative $(\bsx_i,\bsy_i)$,  or its pre-calculated functional approximation coefficients $\widehat\bsbeta_i$ given by \eqref{eq:OLS}.


\section{Experimental Methods}    \label{sec: experiments}

In this section we describe the evaluation of different versions of our method: mixtures of B-spline and polynomial functional regressions with spatial Gaussian gates (resp. SgMFR-Bspl, SgMFR-poly), mixture of B-spline regressions with softmax gates (SsMFR-Bspl), mixture of vectorized B-spline regressions with Gaussian gates (SgMVFR-Bspl).

\subsection{Data}

91 head and neck DECT scans were evaluated, consisting of head and neck squamous cell carcinoma (HNSCC) tumors of different sizes and stages from different primary sites.
34\% of tumors in our dataset are located in oral cavity, 26\% in oropharynx, 21\% in larynx, 8\% in hypopharynx and 11\% in other locations. The tumors T-stage~\cite{TstageManual} ranges from T1 to T4. 75\% of patients were coming for a first diagnostic while 25\% were recurrent patients.
Institutional review board approval was obtained for this study with waiver of informed consent. Tumors were contoured by an expert head and neck radiologist.
All scans were acquired using a fast kVp switching DECT scanner (GE Healthcare) after administration of IV contrast and reconstructed into 1.25mm sections of axial slices with a resolution of 0.61mm, as previously described~\cite{intro22}.
Multienergy VMIs were reconstructed at energy levels from 40 to 140 keV in 5 keV increments at the GE Advantage workstation (4.6; GE Healthcare).

In each DECT scan, we crop Volumes of Interest (VOIs) of size 150*150*6 containing a tumor, along with the 21-point-spectral curve associated to each selected voxel, in order to reduce the computational demands for an exploratory study, and to exclude regions containing a majority of air voxels around the body. A pre-processing step is also applied to mask any remaining air voxels in the VOI to focus the clustering on tissues.

\subsection{Regularization parameter}

In our study, we augmented the statistical estimator in equation \eqref{eq:MLE Rk} of the covariance matrix of the spatial-coordinates within cluster $k$, with a regularization parameter $\lambda \in (0,1]$, which  controls the amount of spatial dispersion (neighborhood) taken into account in the spatial mixture weights, by 
\begin{equation}
  \label{eq:updatedMLERk}
\widetilde\bR_{k}^{(t+1)} = \lambda \bR_{k}^{(t+1)}.
\end{equation}By doing so, we can numerically control the amount of data within cluster $k$ (i.e. its volume). Indeed, if we decompose $\bC_k = \lambda\bD \bA\bD^T$ where $\bA$ is the $\bC_k$'s  eigenvectors matrix, and $\bD$ a diagonal matrix whose diagonal elements are the eigenvalues in decreasing order, then $\lambda$ is the volume of cluster $k$.
Since the tumor cluster has in general no strong spatial dispersion, then in practice, we take small values of order $0.1$ (see examples in~Fig.~\ref{fig:lambda-tuning}).

\subsection{Parameter initialization}

For the sake of reproducibility, we start by initializing the regression mixture and weight parameters of the EM algorithm with a coarse clustering solution given by a Voronoi diagram. We build up Voronoi tiles over the selected voxels in the VOI (a square region where air voxels are deleted) with k-means algorithm applied only on spatial coordinates. To avoid any randomness in the workflow, we start k-means with fixed, regularly spaced, cluster centers, and in the rare cases when EM does not converge following this initialization, we can re-start k-means with random cluster centers.

Then to fix $K$, the parameter defining the number of clusters, and to fix $\lambda$, the spatio-spectral hyper-parameter, we assess on a small training set (10 patients) the three metrics described in Section~\ref{sec: metrics}.
Because computational time can be long, a full grid search would be too expensive, so an iterative process of fixing K and varying $\lambda$, then fixing $\lambda$ and varying K, then fixing again K and varying $\lambda$ is applied.
The process is run similarly for both methods of mixtures of functional regressions and mixtures of vectorized functional regressions. The range of search values is adapted for each method, and 5 search values are taken in each range. In the end, we use the mean of optimal values over the patients in the train set.

\subsection{Baselines}
\label{sec:baseline}

We compare the quantitative and qualitative performance of our methodology with three baseline algorithms. First, we implement a Gaussian Mixture Model (GMM) with the iterative EM algorithm, using the standard non-reshaped algorithm~\cite{McLachlan2000FMM} to cluster the spectral curves, thus leading to not include spatial coordinates. Because several clusters can become empty through the optimization, we fix an initial number of clusters ($K=150$) that ends up in providing, in average, the same resulting number of clusters than in our method (i.e. $K=40$). Second, we implement k-means clustering, using all vector information available, that is, the input vector is build with spectral information (i.e. the energy decay curve points) concatenated to a vector of spatial information (i.e. the 3D coordinates). The number of clusters is picked to be also $K=40$, the number of clusters being stable throughout the optimization. We use the Matlab k-means implementation for images with a reproducible initialization through the built-in `imsegkmeans' function. Third and last, we implement Selective Search, a machine learning graph-based segmentation method for object recognition. Using a region merging hierarchical approach with an SVM classifier to select the hierarchical rank of the resulting regions, the authors published an open-source code \cite{SelSearch}. We apply Selective Search on a low, an intermediate and a high energy level (resp. 40, 65 and 140keV). These energy levels are used instead of the three RGB image channels. We note that Selective Search does not predetermine the number of clusters but specifies a cluster minimum size or favors smaller or larger cluster sizes.

\subsection{Metrics} \label{sec: metrics}

Our clustering methods, as well as the baseline clustering algorithms are all evaluated using the following three metrics:
\begin{enumerate}
    \item a cluster separation index: \textit{Davies-Bouldin index (DB),} computed on spatial content and on spectral content,
    \item a clustering separation index focused only on the relationship between tumor clusters and other clusters: \textit{Davies-Bouldin index on tumor (DB$_{t}$),} an adaptation of DB, computed on spatial and on spectral content,
    \item a segmentation score computed on tumor clusters versus ground truth region: \textit{the Dice similarity score,} that can be computed only on spatial content.
\end{enumerate}

To define tumor clusters, we select the cluster(s) which best cover the tumor, i.e. the ones that give the best Dice score when being merged together.
Several clusters segmenting the tumor area can indeed represent different tumor subparts, but we only know the tumor primary site contour as ground truth.

When $C$ is the ensemble of clusters, $d(\cdot,\cdot)$ is the Euclidean distance operator, $\overline{c_k}$ is the centroid of cluster $c_k$, the Davies-Bouldin index is defined as:
{\small
\begin{equation}
 \!\!\!\!   \text{DB}(C) \!=\! 
    1/|C|
    \textstyle
    \sum_{c_k \in C} \max_{c_l \in C \backslash c_k} 
    (
    S(c_k) + S(c_l)
    )
    \slash
    d(\overline{c_k},\overline{c_l})
\end{equation}
}where
{\small $S(c_k) = 1/|c_k| \sum_{x_i \in c_k} d(x_i,\overline{c_k}).$ } 
\\
The `tumor' Davies-Bouldin index is adapted as:
{\small
\begin{equation}
 \!\!\!   \text{DB}_{t}(C) \!=\! \textstyle\max_{c_l \in C \backslash c_{tum}} (S(c_{tum}) + S(c_l))/d(\overline{c_{tum}},\overline{c_l})),
\end{equation}
}where $c_{tum}$ is the region of the merged tumor clusters. 
\\
The Dice score is defined as:
{\small
\begin{equation}
\text{Dice}(c_{tum},c_{truth}) = 2* |c_{tum} \cap c_{truth}|/(|c_{tum}| + |c_{truth}|)\,,    \end{equation}
}where $c_{truth}$ is the ground truth region.
 We summarize the distribution of these index values across the population by computing the mean, median and interquartile range.

\section{Results} \label{sec: quant-results}

\begin{figure*}[!ht]
\captionsetup[subfigure]{font=small,aboveskip=1pt,belowskip=2pt}
\begin{center}
\begin{subfigure}[t]{.2\textwidth}
  \centering
  \includegraphics[trim={60pt 45 60 20},clip, width=\linewidth]{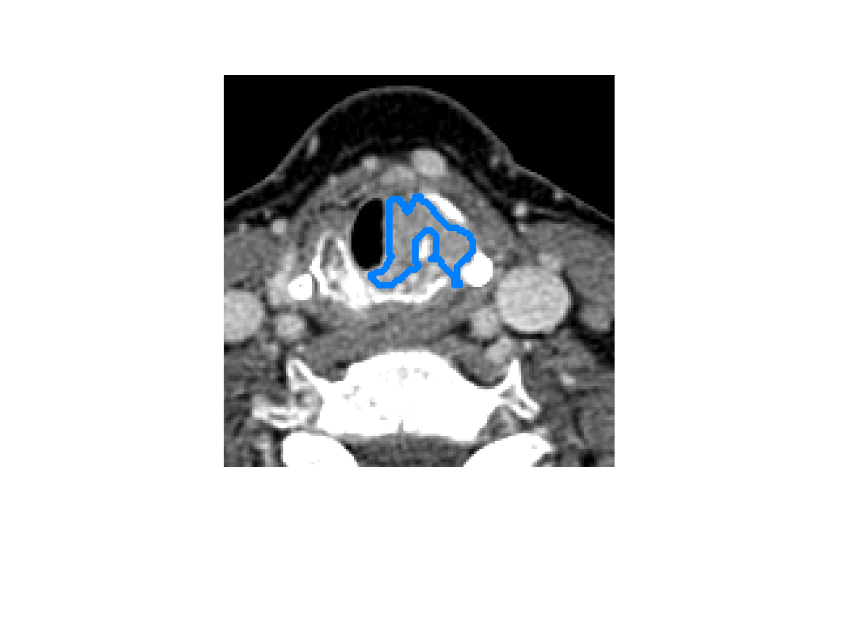}  
  \caption{\textbf{Original slice}}
\end{subfigure}
\begin{subfigure}[t]{.2\textwidth}
  \centering
  \includegraphics[trim={60pt 45 60 20},clip, width=\linewidth]{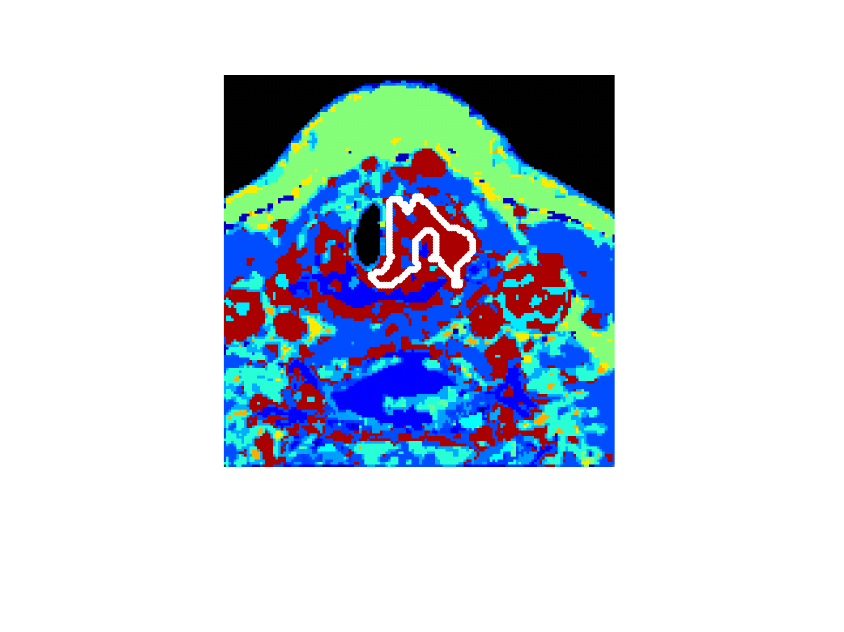} 
  \caption{\textbf{GMM} \\
  Dice=0.221,
  DB=3.7$\big/$1.8,\\
  DB$_t$=20.7$\big/$4.2,
  time=101s
}
\end{subfigure}
\begin{subfigure}[t]{.2\textwidth}
  \centering
  \includegraphics[trim={60pt 45 60 20},clip, width=\linewidth]{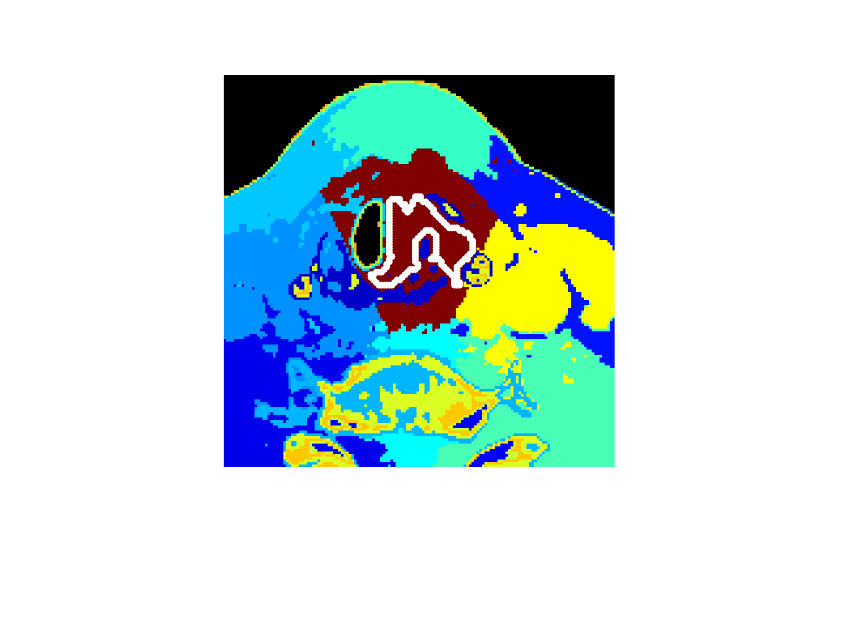} 
  \caption{\textbf{k-means} \\
  Dice=0.409,
  DB=3.8$\big/$15.3,\\
  DB$_t$=2.2$\big/$17.9,
  time=3.47s
}
\end{subfigure}
\begin{subfigure}[t]{.2\textwidth}
  \centering
  \includegraphics[trim={60pt 45 60 20},clip, width=\linewidth]{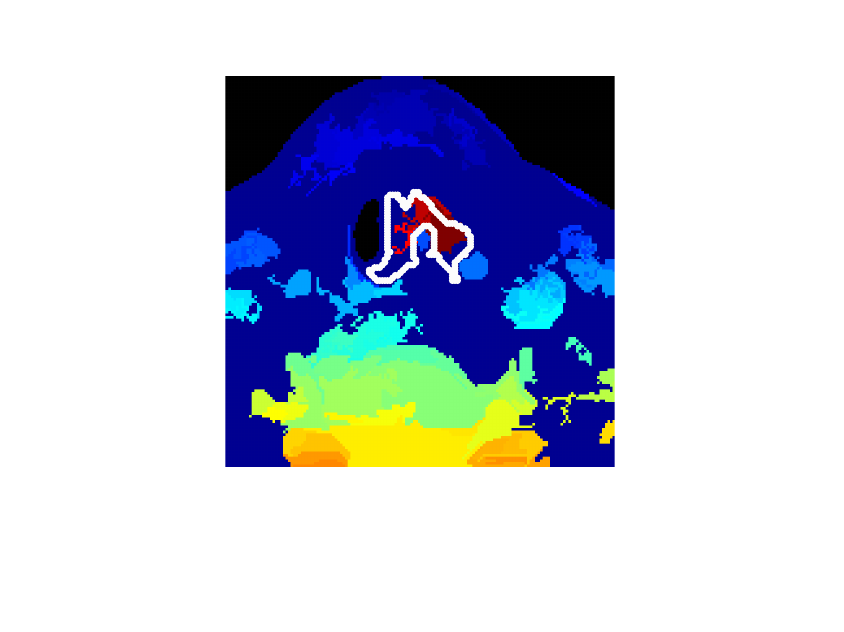} 
  \caption{\textbf{S.Search} \\
Dice=0.523,
DB=1.1$\big/$16.1,\\
DB$_t$=2.7$\big/$15.8,
time=0.08s
}
\end{subfigure}

\hfill

\begin{subfigure}{.2\textwidth}
  \centering
  \includegraphics[trim={60pt 45 60 20},clip, width=\linewidth]{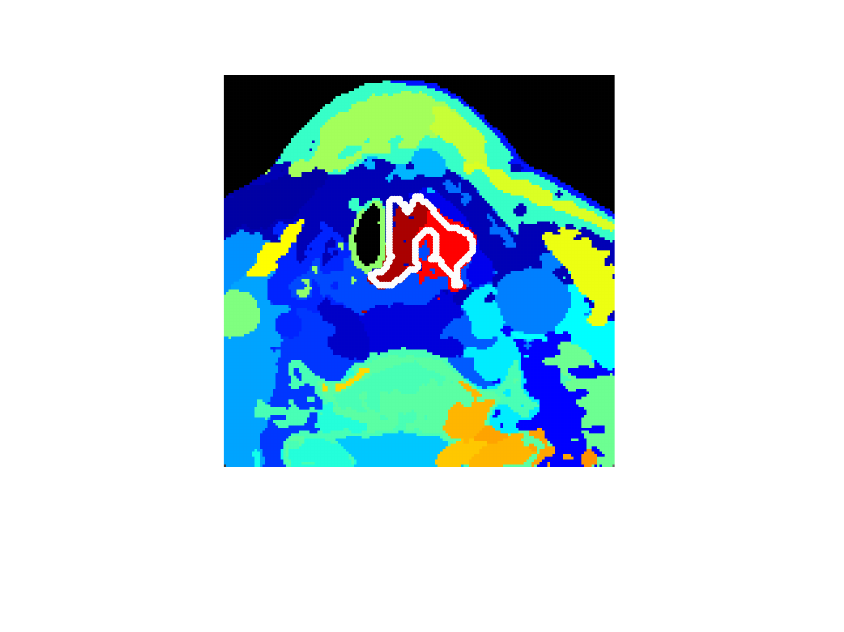}  
  \caption{\textbf{SgMFR-Bspl} \\
  Dice = 0.809,
DB=1.4$\big/$7.6,\\
DB$_t$=4.3$\big/$5.6,
time=1521s
}
\end{subfigure}
\begin{subfigure}{.2\textwidth}
  \centering
  \includegraphics[trim={60pt 45 60 20},clip, width=\linewidth]{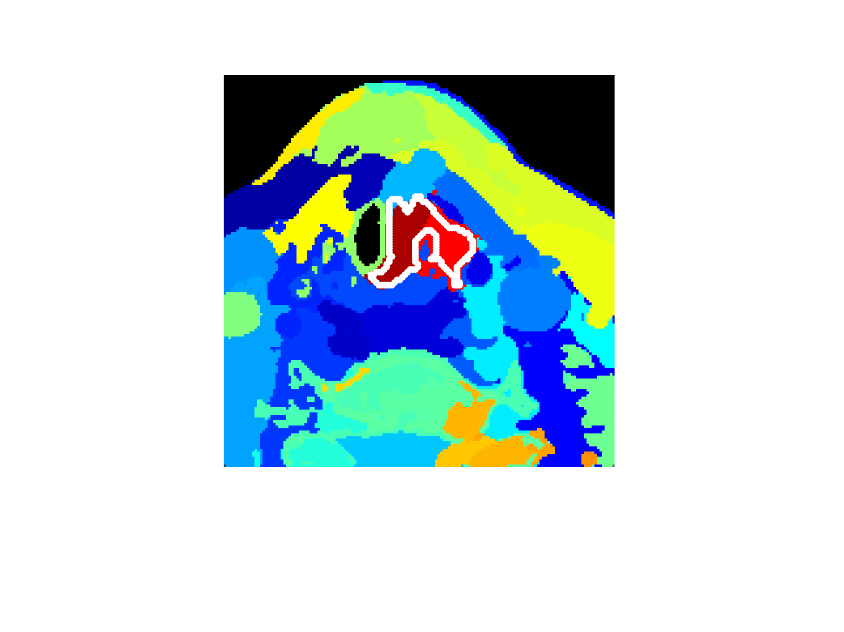} 
  \caption{\textbf{SgMFR-poly} \\
  Dice=0.838,
DB=1.2$\big/$6.5,\\
DB$_t$=1.8$\big/$3.6,
time=603s
}
\end{subfigure}
\begin{subfigure}{.2\textwidth}
  \centering
  \includegraphics[trim={60pt 45 60 20},clip, width=\linewidth]{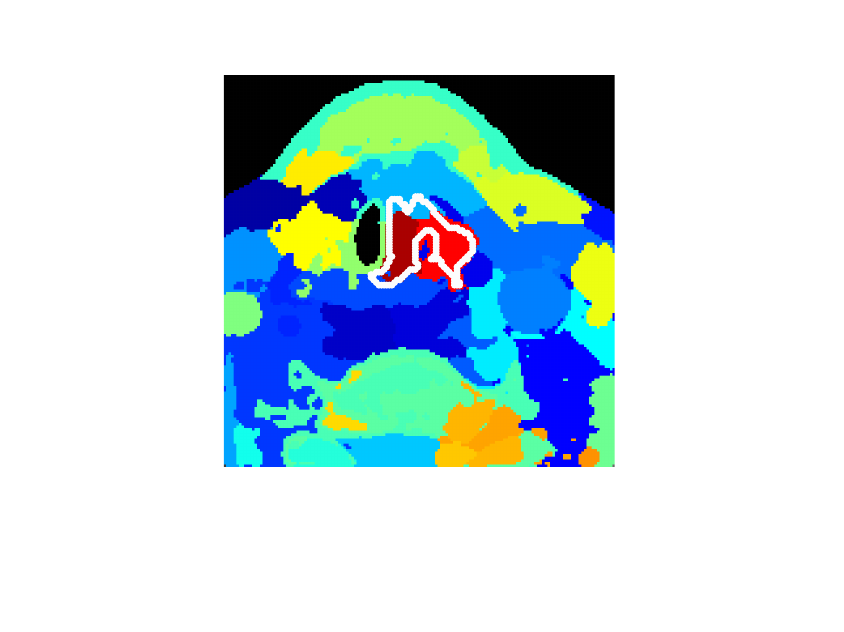} 
  \caption{\textbf{SgMVFR-Bspl} \\
  Dice=0.761,
DB=1.3$\big/$9.0,\\
DB$_t$=2.0$\big/$6.2,
time=199s
}
\end{subfigure}
\begin{subfigure}{.2\textwidth}
  \centering
  \includegraphics[trim={60pt 45 60 20},clip, width=\linewidth]{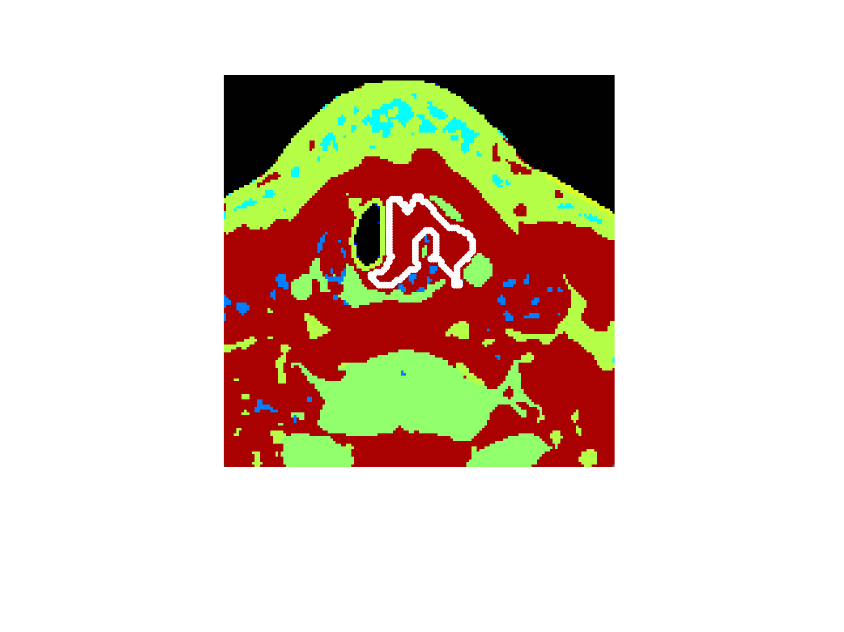} 
  \caption{\textbf{SsMFR-Bspl} \\
Dice=0.109,
DB=0.8$\big/$11.6,\\
DB$_t$=4.0$\big/$0.7,
time=3920s
}
\end{subfigure}

\caption{Qualitative clustering results for each approach in one tumor (DB$_{(t)} =$ spatial/spectral index). Our proposed approaches are on the bottom row. \textit{One random color is assigned per cluster, ground truth tumor contour is in blue (a) or white (b,c,d,e,f,g,h).}}
\label{fig:compar-clust}
\end{center}
\end{figure*}

Fig.~\ref{fig:compar-clust} shows a qualitative overview of our results for one tumor example. We visualize the results on a 2D slice when the model has been run on the 3D VOI containing this slice. The top row shows the performance of the baseline algorithms, whereas the bottom row shows our proposed methods. While baseline approaches attribute a high number of clusters to bone regions containing big spectral variations and miss smaller variations in tissue regions, our methods with Gaussian gates in Fig.~\ref{fig:compar-clust}~(e, f, g) are able to adapt to relative variations and split the image with more spatial coherence. The results also demonstrate that our method is able to capture tissue characteristics invisible in Fig.~\ref{fig:compar-clust}~(b, c, d, h). Note that \text{DB} and \text{DB$_{t}$} scores depend on the number of clusters and SsMFR in Fig.~\ref{fig:compar-clust}~(h) has a very low number of clusters (softmax having vanishing clusters in the optimization). GMM and Selective Search in Fig.~\ref{fig:compar-clust}~(b) and (d) have around 40 clusters (varying number as explained in Section~\ref{sec:baseline}). The results in Fig.~\ref{fig:compar-clust}~(c, e, f, g) were obtained with 40 clusters.

\begin{figure*}[!ht]
\centering
  \includegraphics[trim={140pt 0 120 0},clip, width=\linewidth]{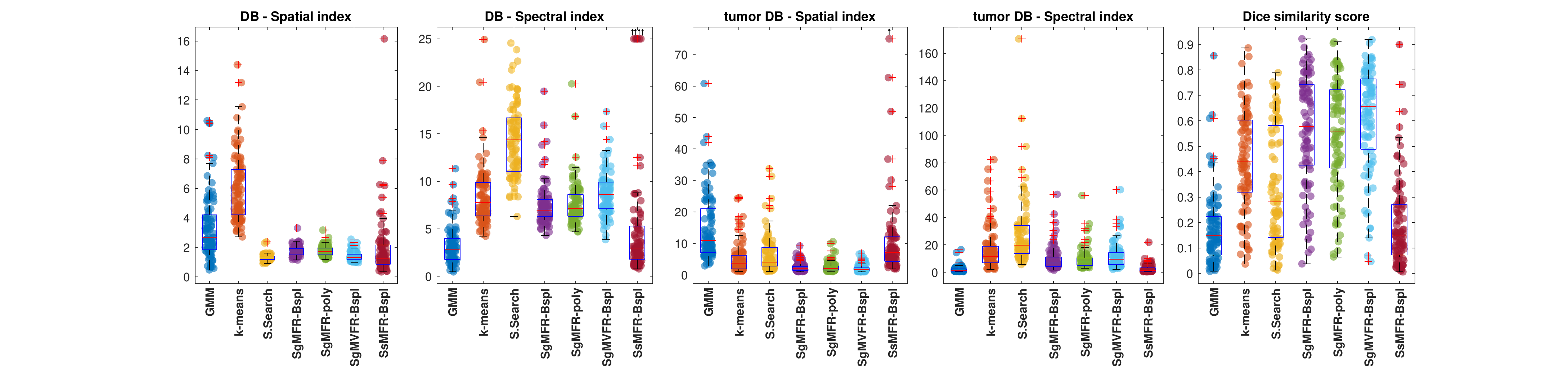} 
  \caption{\small{Boxplots for each metric per method.} \textit{Note that SsMFR-Bspl method gives few outliers out of reach (order of $10^{13}$) on the DB spectral index, and one outlier of 135 for tumor DB spatial index. These values are shifted in the displayed range and exhibit a top arrow.}}
  \label{fig:boxplots}
\end{figure*}

Table~\ref{table:compar-clustering} and Fig.~\ref{fig:boxplots} present the quantitative results obtained with the three clustering metrics defined in Section~\ref{sec: metrics}. Among the three proposed methods that outperform the baseline methods in terms of Dice score (SgMFR-Bspl, SgMFR-poly, and SgMVFR-Bspl), we compared the Dice score distribution obtained with SgMFR-poly (which has the lowest median Dice score among the three) to that obtained wtih the k-means-like baseline (which has the highest median Dice score) with a two-sample t-test, obtaining p=0.0014.

\begin{table*}[!ht]
\begin{center}
\caption{\small{Average, median \textit{(interquartile range)} of metrics and runtime over 81 patient scans. \textit{DB cluster separation index is computed for spatial content (spat-DB) and spectral content (spec-DB); idem for DB$_t$ indices focused on tumor separability; runtime is given in seconds.}}}
\resizebox{\textwidth}{!}{
\setlength{\tabcolsep}{3pt}
\begin{tabular}{l|ccc|ccc|ccc|ccc|ccc|ccc|ccc}
\toprule
& \multicolumn{3}{c}{GMM}
& \multicolumn{3}{|c}{k-means}
& \multicolumn{3}{|c}{S.Search}
& \multicolumn{3}{|c}{SgMFR-Bspl}
& \multicolumn{3}{|c}{SgMFR-poly}
& \multicolumn{3}{|c}{SgMVFR-Bspl}
& \multicolumn{3}{|c}{SsMFR-Bspl}
\\
\midrule
& avg & med & itq
& avg & med & itq
& avg & med & itq
& avg & med & itq
& avg & med & itq
& avg & med & itq
& avg & med & itq
\\
\midrule
Dice
& 0.17 & 0.15 & \textit{(0.15)}
& 0.45 & 0.44 & \textit{(0.28)}
& 0.35 & 0.28 & \textit{(0.44)}
& 0.56 & 0.58 & \textit{(0.32)}
& 0.57 & 0.56 & \textit{(0.31)}
& 0.61 & 0.66 & \textit{(0.28)}
& 0.20 & 0.15 & \textit{(0.20)}
\\
spat-DB
& 3.32 & 2.69 & \textit{(2.38)}
& 6.00 & 5.57 & \textit{(3.08)}
& 1.30 & 1.28 & \textit{(0.24)}
& 1.75 & 1.75 & \textit{(0.45)}
& 1.74 & 1.72 & \textit{(0.45)}
& 1.40 & 1.32 & \textit{(0.35)}
& 1.84 & 1.22 & \textit{(1.31)}
\\
spec-DB
& 3.17 & 2.80 & \textit{(2.21)}
& 8.41 & 7.75 & \textit{(3.51)}
& 14.36 & 14.36 & \textit{(5.66)}
& 7.52 & 6.95 & \textit{(1.83)}
& 7.66 & 7.13 & \textit{(2.27)}
& 8.69 & 8.58 & \textit{(2.81)}
& 3.78e12 & 2.96 & \textit{(3.48)}
\\
spat-DB$_t$
& 14.97 & 10.95 & \textit{(13.94)}
& 5.07 & 3.68 & \textit{(4.30)}
& 6.57 & 4.07 & \textit{(6.00)}
& 2.49 & 2.05 & \textit{(1.27)}
& 2.59 & 2.00 & \textit{(1.33)}
& 2.05 & 1.70 & \textit{(0.95)}
& 11.07 & 7.12 & \textit{(7.89)}
\\
spec-DB$_t$
& 1.97 & 1.11 & \textit{(1.71)}
& 16.52 & 11.42 & \textit{(12.17)}
& 28.47 & 19.86 & \textit{(20.11)}
& 9.64 & 6.78 & \textit{(6.90)}
& 9.46 & 7.52 & \textit{(5.46)}
& 11.57 & 9.53 & \textit{(8.64)}
& 2.52 & 1.36 & \textit{(2.20)}
\\
runtime
& 342 & 165 & \textit{(210)}
& 3.64 & 3.78 & \textit{(0.78)}
& 0.062 & 0.047 & \textit{(0.014)}
& 1423 & 1321 & \textit{(600)}
& 902 & 901 & \textit{(345)}
& 437 & 361 & \textit{(277)}
& 2425 & 2121 & \textit{(2049)}
\\
\bottomrule
\end{tabular}
}
\label{table:compar-clustering}
\end{center}
\end{table*}

\begin{figure}[!t]
\captionsetup[subfigure]{aboveskip=1pt,belowskip=2pt}
\begin{center}
\begin{subfigure}[t]{.15\textwidth}
  \centering
  \includegraphics[trim={60pt 45 60 20},clip, width=\linewidth]{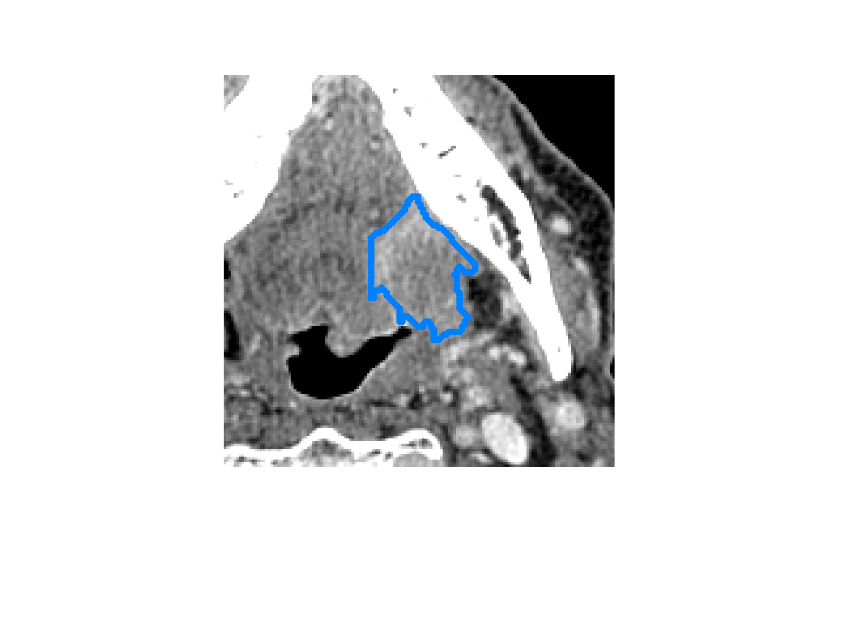}  
  \caption{Original slice}
\end{subfigure}
\begin{subfigure}[t]{.15\textwidth}
  \centering
  \includegraphics[trim={60pt 45 60 20},clip, width=\linewidth]{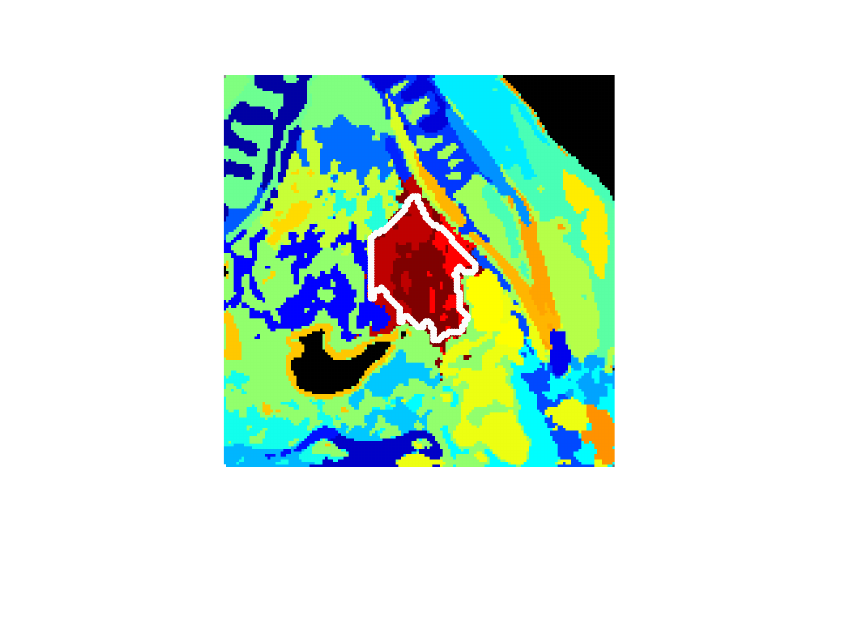} 
  \caption{$\lambda = {\bf 0.075}$,\\
  Dice = 0.79,\\
  DB = 1.68$\big/$7.29,\\
  DB$_t$ = 1.47$\big/$6.79}
\end{subfigure}
\begin{subfigure}[t]{.15\textwidth}
  \centering
  \includegraphics[trim={60pt 45 60 20},clip, width=\linewidth]{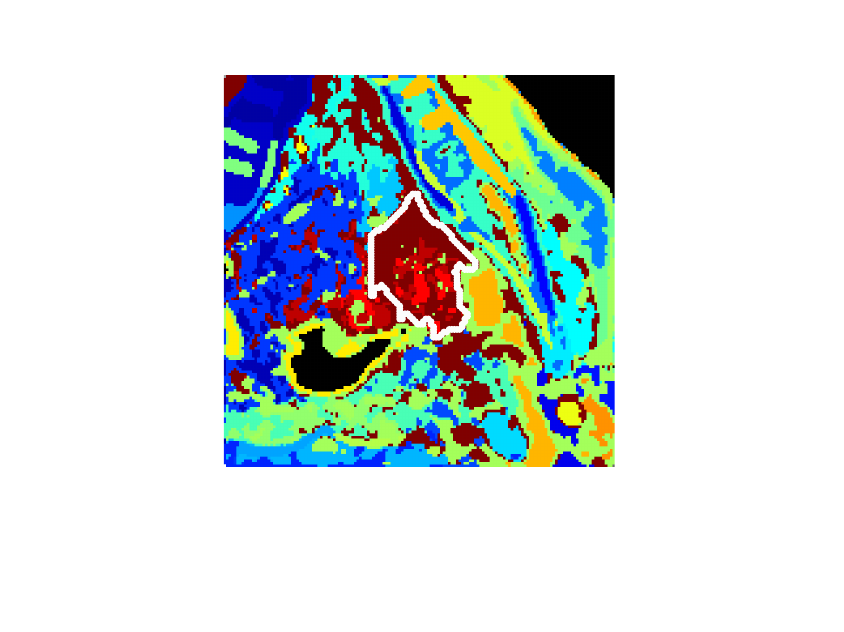} 
  \caption{$\lambda = {\bf 0.100}$,\\
  Dice = 0.39,\\
  DB = 2.59$\big/$4.27,\\
  DB$_t$ = 3.34$\big/$3.74}
\end{subfigure}

\hfill

\begin{subfigure}[t]{.15\textwidth}
  \centering
  \includegraphics[trim={60pt 45 60 20},clip, width=\linewidth]{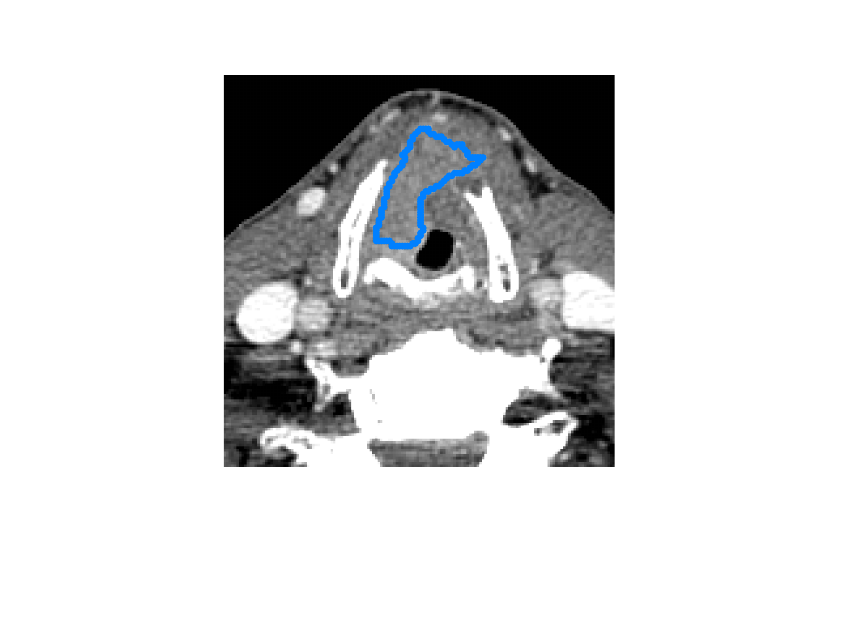}  
  \caption{Original slice}
\end{subfigure}
\begin{subfigure}[t]{.15\textwidth}
  \centering
  \includegraphics[trim={60pt 45 60 20},clip, width=\linewidth]{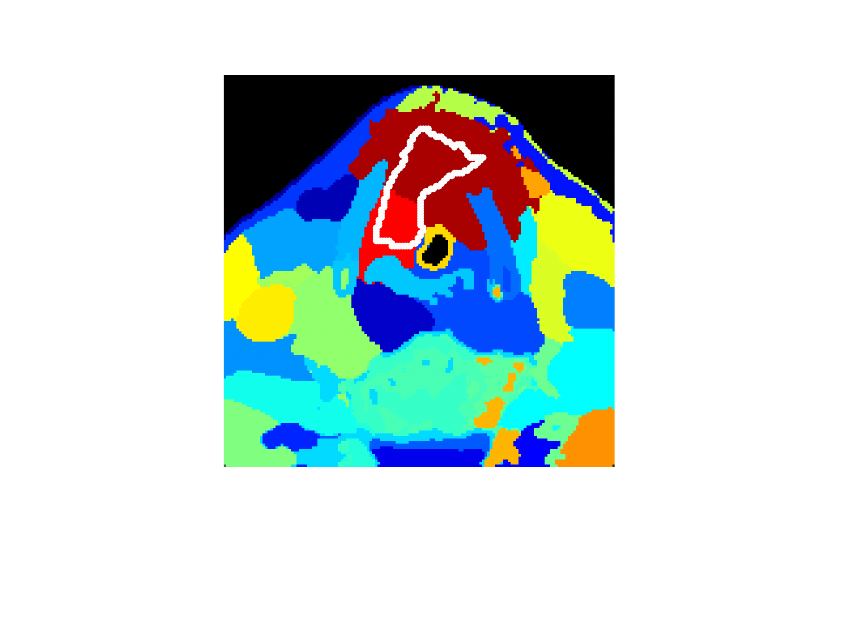} 
  \caption{\textbf{$\lambda = {\bf 0.075}$},\\
  Dice = 0.36,\\
  DB = 1.75$\big/$6.70,\\
  DB$_t$ = 1.28$\big/$23.71}
\end{subfigure}
\begin{subfigure}[t]{.15\textwidth}
  \centering
  \includegraphics[trim={60pt 45 60 20},clip, width=\linewidth]{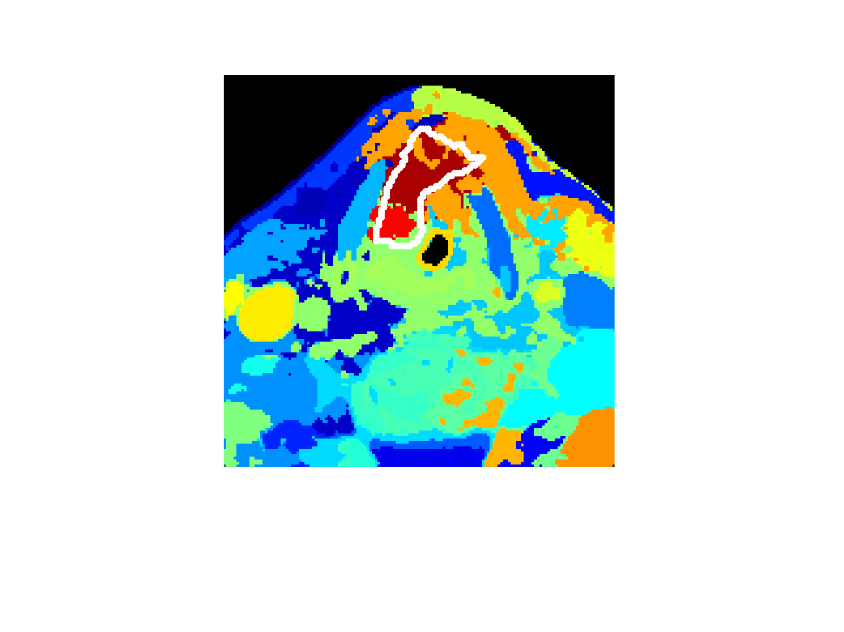} 
  \caption{\textbf{$\lambda = {\bf 0.080}$},\\
  Dice = 0.64,\\
  DB = 1.87$\big/$6.57,\\
  DB$_t$ = 1.78$\big/$2.93}
\end{subfigure}

\caption{Clustering results for our SgMFR method with different $\lambda$ tuning. 
\textit{One random color is assigned per cluster, ground truth tumor contour is in blue (a,d) or white (b,c,e,f).}}
    \label{fig:lambda-tuning}
\end{center}
\end{figure}

\begin{figure}[!t]
\captionsetup[subfigure]{aboveskip=1pt,belowskip=2pt}
\begin{center}
\begin{subfigure}[t]{.22\textwidth}
  \centering
  \includegraphics[trim={60pt 45 60 20},clip, width=\linewidth]{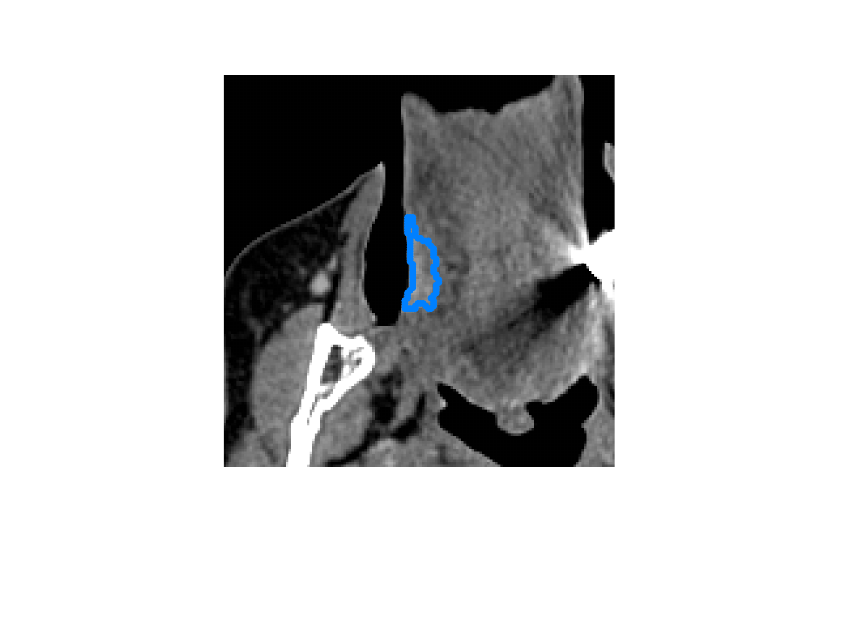}  
  \caption{Original slice}
\end{subfigure}
\begin{subfigure}[t]{.22\textwidth}
  \centering
  \includegraphics[trim={60pt 45 60 20},clip, width=\linewidth]{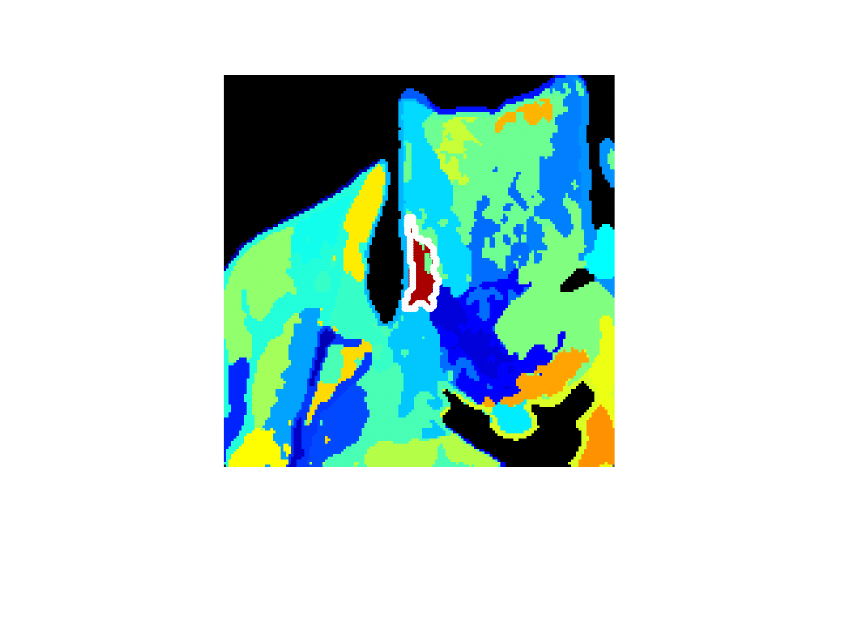} 
  \caption{Dice=0.84, DB=1.64$\big/$6.92,\\
  DB$_t$=1.98$\big/$7.22, $\lambda=0.075$}
\end{subfigure}
\caption{Clustering results with our SgMFR for a small tumor. Note the robustness of the result in the presence of a metallic artifact in the right hand side of the anatomical image. 
}
\label{fig:small-tumor}
\end{center}
\end{figure}

Fig.~\ref{fig:lambda-tuning} showcases the clustering results obtained when varying the tuning of $\lambda$. Here, we understand that a smaller $\lambda$ gives a higher preference to the spatial information: clusters are compact and are defining well-separated areas. On the other hand, a larger $\lambda$ gives a higher priority to the spectral information: clusters more closely match tissue characteristics, but one cluster can be split into tiny voxel groups spread all over the image. The ideal $\lambda$ choice would be a $\lambda$ that prioritizes spectral information, but still achieves some cluster spatial compactness. We assess this through metrics calculated on spectral and spatial content as explained in Section \ref{sec: metrics}. The general tuning of $\lambda = 0.075$ has been determined to be, on average, the optimal hyper-parameter. However, we can see strong improvements in tumor separation, on a case by case basis, with small variation of $\lambda$. As shown in Fig.~\ref{fig:lambda-tuning}~(e) and (f), the Dice score increases from 0.36 to 0.64. This shows an example out of several results belonging to the lower quartile in the boxplot of Dice scores in Fig.~\ref{fig:boxplots} that could be highly improved simply with a specific tuning. Some other examples of results in the lower quartile could be due to small tumors (size inferior to 1cm), although half of these small tumors are actually well separated with our method, reaching a Dice score as high as 0.84 in the best case (see in Fig.~\ref{fig:small-tumor}).

\section{Discussion and Conclusion}    \label{sec: discussion}

In this paper, we have developed a statistical methodology to cluster intensity attenuation curves in DECT scans.
We applied our proposed methods, together with other alternative clustering algorithms used as baselines, to a set of 91 DECT scans of HNSCC tumors. The classical manner of evaluating algorithms for clustering/segmentation is via measures of overlap (such as the Dice score) with a ground truth segmentation. However, as mentioned in the Introduction above, the manual segmentations of HNSCC tumors that are used as ``ground truth'' can suffer from large inter-rater variability, and do not incorporate in any systematic manner regions immediately adjacent to the tumor that may be biologically important for determining the course of evolution of the tumor. Because of this inherent uncertainty in the appropriate contours of an HNSCC tumor, the main objective in our paper was to compare our clustering results to the manual contouring, but also to explore associations between voxels within the ground truth tumor contour and voxels in the surrounding tissue areas.

Compared to the baseline algorithms, it is clear both visually and quantitatively that our methods using Gaussian gates (SgM(V)FR) produce results that match better the manual segmentation contours. Our method using softmax gates (SsMFR) is less flexible compared to the one with Gaussian gating functions, and thus sometimes leads to non-satisfactory results. Although in terms of qualitative assessment clusters of SsMFR-Bspl are indeed more spatially compact, quantitative performance in some situations stays similar to GMM baseline. Thus this variant of the algorithm does not appear to perform well in practice. 
Using Gaussian gates, however, Dice score distributions are significantly better than with the k-means-like algorithm, the best of our baseline methods.

That being said, it is also clear that with Dice scores ranging from nearly 0 to nearly 1, our proposed methods do not recover the ``ground truth'' segmentations in a reliable and consistent manner. Several reasons may explain this finding. First, our clinical dataset of DECT scans is not uniform, i.e. it includes tumors of highly variable characteristics, in highly variable sizes, locations and environments, which makes it particularly challenging. 
Moreover, as seen in Fig.~\ref{fig:lambda-tuning}, changes in parameter tuning can lead to substantial improvement in Dice scores for some tumors. 
Finally, because of their intricate morphology and often small sizes, HNSCC tumors are inherently difficult to segment. In a recent international challenge, Dice scores of head and neck tumor segmentation ranged across different competition entries between 0.56 and 0.76 \cite{HECKTOR}.

Most importantly, as argued throughout this paper, the clinical value of recovering the manual segmentations of HNSCC tumors as an objective criterion for evaluating the algorithm is also not clear. In fact, it has recently been argued in the clinical literature that AI methods in medical imaging would be more meaningful if evaluated against clinical outcomes, as opposed to an evaluation against radiologists' performance, due to inherent subjectivity and variability of the latter~\cite{PeterDemystification2019}. For all the reasons, the objective of this paper was moved away from reproducing the manual contours produced by the radiologist, and was focused instead on developing tools that discover patterns of association in the DECT data. 

Our study has several limitations. As discussed above, in our view, the appropriate way of evaluating the methodology's clinical utility is not by computing Dice scores relative to manually drawn contours. Rather, a more clinically informative evaluation would determine the performance of the recovered clusters in predicting clinical outcome in a machine learning setting, compared to the same predictive algorithm applied with the manual tumor segmentations. Such an evaluation is missing from the present paper; it will be part of a subsequent paper in future work. Another limitation stems from the lack of an automated identification of those clusters that are associated with the tumor region. Right now, we choose those clusters that maximize overlap with the manually segmented tumor region. Ideally, however, the abnormal tumor clusters should be identified automatically, by selecting those clusters that have the highest association with clinical outcomes. In this manner, the automated cluster identification can be naturally made part of a single machine learning pipeline for predicting clinical outcomes. Yet another limitation comes from the very small size of the subset of tumors (N=10) over which we estimated the algorithm parameters ($\lambda$ and number of clusters), before applying the algorithm to the remaining 81 tumors in our dataset. A larger  dataset, together with additional patient-specific  tuning will help tune the algorithm's performance.

The need for the improvements described above is clear, and they will be made part of a subsequent publication. In the present article, we chose to focus on the theoretical and algorithmic developments. As mentioned in the Introduction, this is the first time to our knowledge that statistical tools from the Functional Data Analysis field are put into practice with DECT data. As such, the present paper remains an inherently exploratory one in its experimental framework.

Nevertheless, we believe we provide several important technical and methodological contributions.
We constructed a functional regression mixture model that integrates spatial content into the mixture weights, and we developed a dedicated EM algorithm to estimate the optimal model parameters. Our mixture-based model is a highly flexible statistical approach allowing for many choices of the parametric form of the component densities.  We proposed two candidate designs for the mixture weights, normalized Gaussian gates and softmax gates. The Gaussian-gate closed-form solution for spatial mixture weight updates considerably reduces the computation time while also providing solutions with better clustering index values, compared to the Newton-Raphson optimization algorithm needed at each update of the softmax-gating parameters.






\bibliographystyle{IEEEtran}  
{
\bibliography{refs}
}

\end{document}